\documentclass[conference]{IEEEtran}
\IEEEoverridecommandlockouts
\usepackage{cite}
\usepackage{amsmath,amssymb,amsfonts}
\usepackage{algorithmic}
\usepackage{graphicx}
\usepackage{textcomp}
\usepackage{xcolor}
\usepackage{float}
\usepackage{array}
\usepackage{xurl}
\usepackage{booktabs}
\def\BibTeX{{\rm B\kern-.05em{\sc i\kern-.025em b}\kern-.08em
    T\kern-.1667em\lower.7ex\hbox{E}\kern-.125emX}}
\def\BibTeX{{\rm B\kern-.05em{\sc i\kern-.025em b}\kern-.08em
    T\kern-.1667em\lower.7ex\hbox{E}\kern-.125emX}}

\begin{document}
\newpage
\thispagestyle{empty}
\onecolumn

\begin{center}
{\Huge \textbf{IEEE Copyright Notice}}\\[2em]

\parbox{0.95\linewidth}{
\Large
© 2025 IEEE. Personal use of this material is permitted. Permission from IEEE must be obtained for all 
other uses, in any current or future media, including reprinting/republishing this material for advertising 
or promotional purposes, creating new collective works, for resale or redistribution to servers or lists, 
or reuse of any copyrighted component of this work in other works.
}

\vspace{3em}

{\Large
Accepted for publication at the \textit{IEEE PES Transmission and Distribution Conference and Exposition (T\&D) 2026},
May 4--7, 2026.
}

\vspace{2em}

\end{center}

\newpage
\twocolumn

\title{Understanding Regional Inertia Dynamics in CAISO from Real Grid Disturbances\\

\thanks{This manuscript has been authored by UT-Battelle, LLC, 
under contract DE-AC05-00OR22725 with the US Department 
of Energy (DOE). The US government retains and the 
publisher, by accepting the article for publication, 
acknowledges that the US government retains a nonexclusive, 
paid-up, irrevocable, worldwide license to publish or reproduce 
the published form of this manuscript, or allow others to do so, 
for US government purposes. DOE will provide public access 
to these results of federally sponsored research in accordance 
with the DOE Public Access Plan 
(https://www.energy.gov/doe-public-access-plan).
 
}
}

\author{
    \IEEEauthorblockN{Saurav Dulal\textsuperscript{1},  Mohammed M. Olama\textsuperscript{2}, Ali R. Ekti\textsuperscript{2}, Nils M. Stenvig\textsuperscript{2}, and Yilu Liu\textsuperscript{1,2}
    }
    \IEEEauthorblockA{
    \textsuperscript{1} Department of Electrical Engineering and Computer Science \\
    The University of Tennessee, Knoxville, TN, USA \\
    \textsuperscript{2} Oak Ridge National Laboratory, Oak Ridge, TN, USA
    }
    (emails: sdulal@vols.utk.edu, olamahussemm@ornl.gov, ektia@ornl.gov, stenvignm@ornl.gov, liu@utk.edu)
}

\maketitle

\begin{abstract}
The shift from synchronous generators to inverter-based resources has caused power system inertia to be unevenly distributed across power grids. As a result, certain grid regions are more vulnerable to high rate-of-change of frequency (RoCoF) during disturbances. This paper presents a measurement-based framework for estimating grid inertia in CAISO (California Independent System Operator) region using real disturbance-driven frequency data from the Frequency Monitoring Network (FNET/GridEye). By analyzing confirmed disturbances from 2013 to 2024, we identify trends in regional inertia and frequency dynamics, highlighting their relationship with renewable generation and the evolving duck curve. Regional RoCoF values were up to six times higher than interconnection-wide values, coinciding with declining inertia. Recent recovery in inertia is attributed to the increased deployment of battery energy storage systems with synthetic inertia capabilities. These findings underscore the importance of regional inertia monitoring, strategic resource planning, and adaptive operational practices to ensure grid reliability amid growing renewable integration.

\end{abstract}


\section{Introduction}
The evolving mix of energy resources in power systems has brought significant challenges to frequency stability. Regions with a high share of inverter-based resources (IBRs) are experiencing increasingly reduced inertia, making them more vulnerable to frequency excursions during disturbances. This issue is particularly evident in CAISO (California Independent System Operator), where solar deployment is among the highest in North America, and midday net generation from synchronous generators (SG) regularly dips below nighttime values \cite{EIAduckcurve}.

The impact of low inertia is no longer theoretical. Recent regional and inter-regional blackouts have demonstrated how regional rate-of-change of frequency (RoCoF) exceedances can cascade into widespread instability. For example, the April 2025 blackout in the Iberian Peninsula was initiated by a large-generation loss during a period when nearly 60\% of Spain’s demand was supplied by solar power \cite{linnert2025blackout}. The event escalated due to low-inertia conditions and weak dynamic coupling, leading to the separation of the Iberian grid from the Continental European system. These incidents highlight the urgent need for tools to assess and monitor inertia at a regional scale, particularly in regions like CAISO.

This challenge is equally relevant in the Western United States (U.S.), where CAISO continues to see rapid growth in renewable generation. It reached a record in April 2024, when renewables briefly supplied over 117\% of the system’s demand \cite{CAISOrenewables}. Such conditions can lead to regional variations in inertia, causing certain areas to experience unsafe RoCoF levels even when the overall interconnection response remains within acceptable limits. This disconnect poses critical risks for real-time operations and long-term planning. As such, enhancing situational awareness of regional inertia is essential to ensure reliable grid performance under evolving resource mixes.

Numerous methods have been applied in the literature to estimate system inertia. Model-based approaches such as generator summation rely on dispatch records or known machine parameters \cite{du2023renewable}, but they fail to account for synthetic or load-based inertia contributions. Ambient data methods infer inertia from naturally occurring load fluctuations, though they often struggle with measurement noise  \cite{inertiaWECC}. Probing signal techniques \cite{10252847} offer improved control and observability, but their deployment is intrusive and costly. Event-based estimation methods \cite{dulal2024inertia}, those that leverage frequency response data from actual disturbances, have gained traction as a practical and accurate alternative. These event-based methods capture how the system responds under stress, including contributions from all sources of inertial response, whether synchronous or non- synchronous.

This paper contributes a practical and scalable methodology for estimating regional inertia using real-world frequency data from  Frequency Monitoring Network (FNET/GridEye). Unlike prior studies \cite{inertiaWECC,dulal2024inertia} that focused on interconnection-wide study, our approach explicitly focuses on regional dynamics. We introduce a RoCoF estimation technique based on a sliding time window and apply it to calculate regional inertia using the swing equation. In addition, we develop and apply multiple regional inertia metrics that include local and interconnection-level comparisons and inertial support arrival time to assess spatial variability in inertia. This methodology is performed across the confirmed CAISO disturbances (2013–2024) that offers insights into evolving inertia trends and the influence of IBRs.

The remainder of this paper is organized as follows: Section
II explains the employed frequency data and CAISO frequency dynamics; Section III defines the regional inertia and outlines the measurement framework and estimation methodology; Section IV
presents the results across multiple events and their analysis; and
Section V concludes the paper.

\section{Data Overview and CAISO Frequency Dynamics}
For this study, we used the frequency disturbance recorders (FDRs) that are deployed throughout the U.S. power grid via the FNET/GridEye as illustrated in Fig. 1 \cite{FNET}. These GPS synchronized FDR units measure the frequency, voltage, and the phase angle from different locations of the U.S. interconnections. The measured data is continuously sent to the data server situated at the University of Tennessee, Knoxville (UTK) and Oak Ridge National Laboratory (ORNL). In particular, we need two types of data for this study. These include the frequency response data during events and power imbalance values that are confirmed by the North American Electric Reliability Corporation (NERC).

When a large disturbance occurs in a power system, frequency measurements across different regions exhibit distinct temporal responses. Areas located closer to the disturbance tend to experience a quicker and sharper frequency deviation, while those farther away observe a more delayed and attenuated response. Additionally, the onset time and RoCoF in each region are affected by both the electrical coupling strength and the underlying network topology within the interconnection.
This phenomena is illustrated in Fig. 2, where FDRs located in CA (highlighted in yellow) measure a steep and rapid frequency drop immediately after the event. In contrast, FDRs from other states, such as WA, CO, AZ, UT, NM, MT, and ID (shown in gray), exhibit a more gradual decline in frequency. This temporal divergence demonstrates how inertial support from electrically distant parts of the interconnection arrives with a delay, resulting in a sharp initial RoCoF in CAISO during the disturbance.

This highlights the importance of regional inertia estimation for CAISO, where high penetration of IBRs and lower synchronous inertia can exacerbate the severity of frequency excursions during grid disturbances. Understanding these dynamics is critical for designing effective frequency support strategies and ensuring grid stability in low-inertia conditions.

\begin{figure}[h]
\centering
\includegraphics[width=\linewidth]{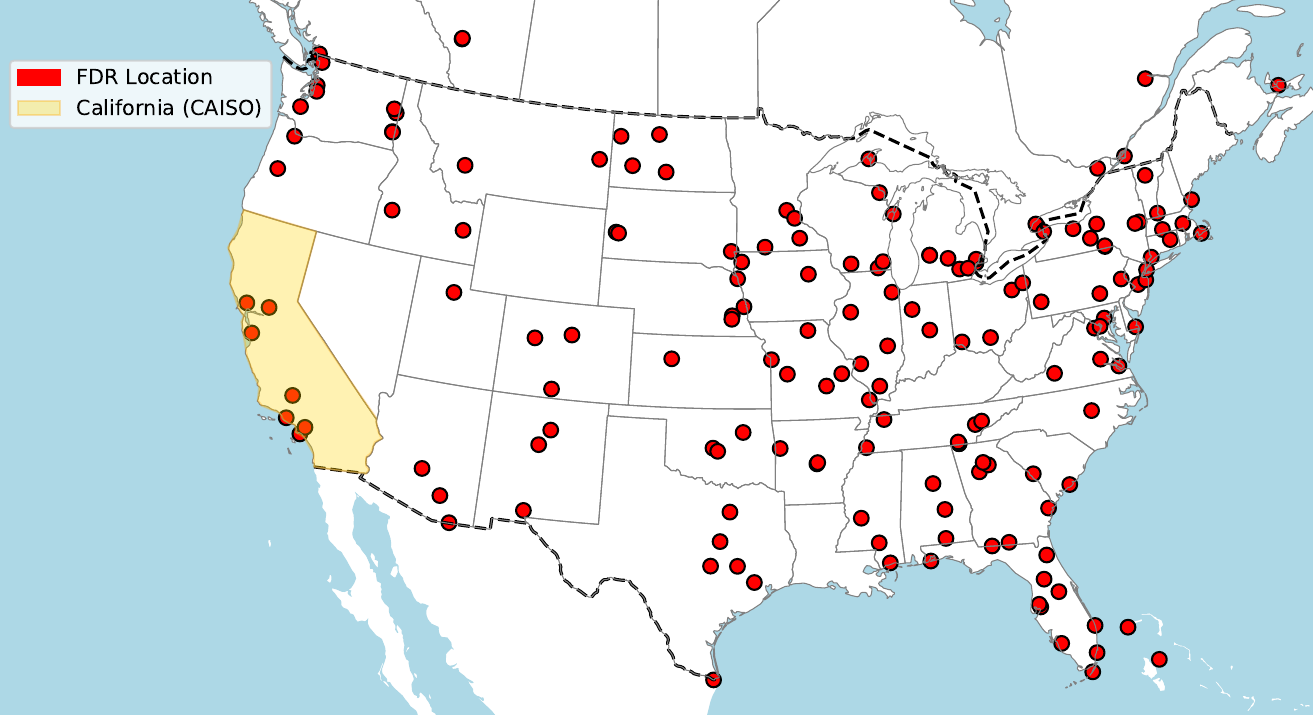}
\caption{Geographical distribution of FDRs across the U.S. with the CAISO region highlighted for regional inertia analysis.}
\label{fig:FDR map}
\end{figure}

\begin{figure}[h]
\centering
\includegraphics[width=\linewidth]{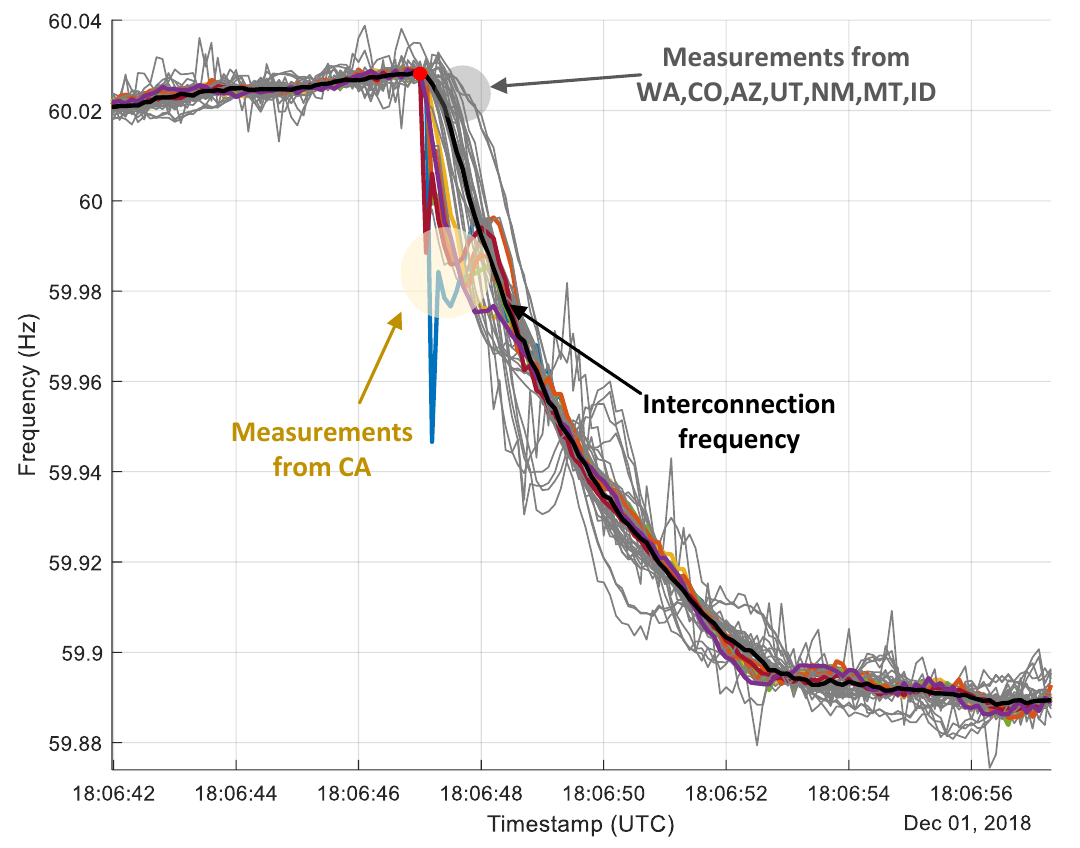}
\caption{Frequency response of the U.S. Western Interconnection during a generation trip in CAISO.}
\label{fig:CAISO dynamics}
\end{figure}

\section{Regional Inertia Estimation Concept and Framework}
The frequency response during large disturbances, particularly the sharp initial response observed in CAISO, emphasizes that inertia is dynamic and regionally dependent. Unlike traditional definitions rooted in generator design specifications, we adopt an empirical view of inertia based on how a region physically reacts to a sudden power imbalance.
In this study, regional inertia is defined as the ability of a localized portion of the grid to resist frequency acceleration immediately following a disturbance. It is not inferred from static system models, but extracted directly from measurements of frequency and power imbalance during real-world events.

\begin{figure}[h]
\centering
\vspace{-0.25 cm}
\includegraphics[width=\linewidth]{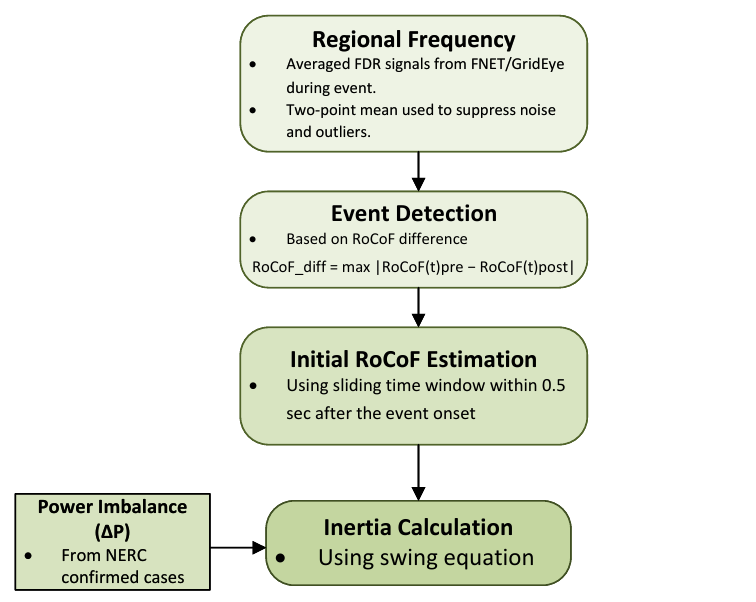}
\caption{Event-based regional inertia estimation process.}
\label{fig:framework}
\end{figure}

To formalize this concept, consider the response of a region to a disturbance with a known power mismatch $\Delta P$. The immediate RoCoF observed within the area, denoted as $\frac{df_{\text{region}}}{dt}$, is used to infer the region’s effective inertia using the following rearranged form of the swing equation \cite{kundur1994chapter3}:

\begin{equation}
H_{\text{region}} S_{\text{region}} = \frac{\Delta P \cdot f_s}{2 \cdot \frac{df_{\text{region}}}{dt}}
\label{eqn:swing}
\end{equation}
where $H_{\text{region}}$ is the regional inertia constant, $S_{\text{region}}$ is the regional power capacity, $f_s$ is the nominal system frequency, $f_{region}$ is the regional frequency and $\Delta P$ is the active power mismatch. The proposed framework for the event-driven regional inertia estimation is illustrated in Fig. \ref{fig:framework} and described as follows:

\subsection{Regional Frequency Estimation}

To estimate the regional frequency, we can compute a weighted average of measurements from multiple FDRs located across CAISO. In the absence of detailed dispatch and inertia allocation data, equal weights are assumed across all FDRs in the region.
To characterize the region’s dynamic response, we compute a representative frequency signal from distributed FDRs. Assuming equal influence across all devices, the regional frequency is:

\begin{equation}
f_{\text{region}} = \frac{1}{n} \sum_{i=1}^{n} f_i
\end{equation}

This spatial averaging inherently suppresses localized fluctuations and captures coherent behavior across the region. However, in scenarios where individual FDRs exhibit pronounced noise, we further apply a two-point mean filter to smooth the frequency. The two-point mean filter averages each data point with its immediate neighbor. It reduces the impact of sudden measurement spikes while preserving the underlying trends.

\subsection{Event Onset Detection}

The event point (onset of frequency deviation) is detected by identifying the timestamp with maximum RoCoF variation across two moving time windows as illustrated in Fig.~\ref{fig:rocof}. In the figure, the red point denotes the event's start point and the slope of the yellow line and the blue lines indicates the pre RoCoF and the post RoCoF. Intuitively, one can observe that the red point provides the maximum difference in the two RoCoFs calculated as:  

\begin{equation}
{RoCoF}_{\text{Diff}} = \max \left| RoCoF_{\text{pre}}(t) - RoCoF_{\text{post}}(t) \right|
\end{equation}

\subsection{RoCoF Computation }

\begin{figure}[h]
\centering
\includegraphics[width=\linewidth]{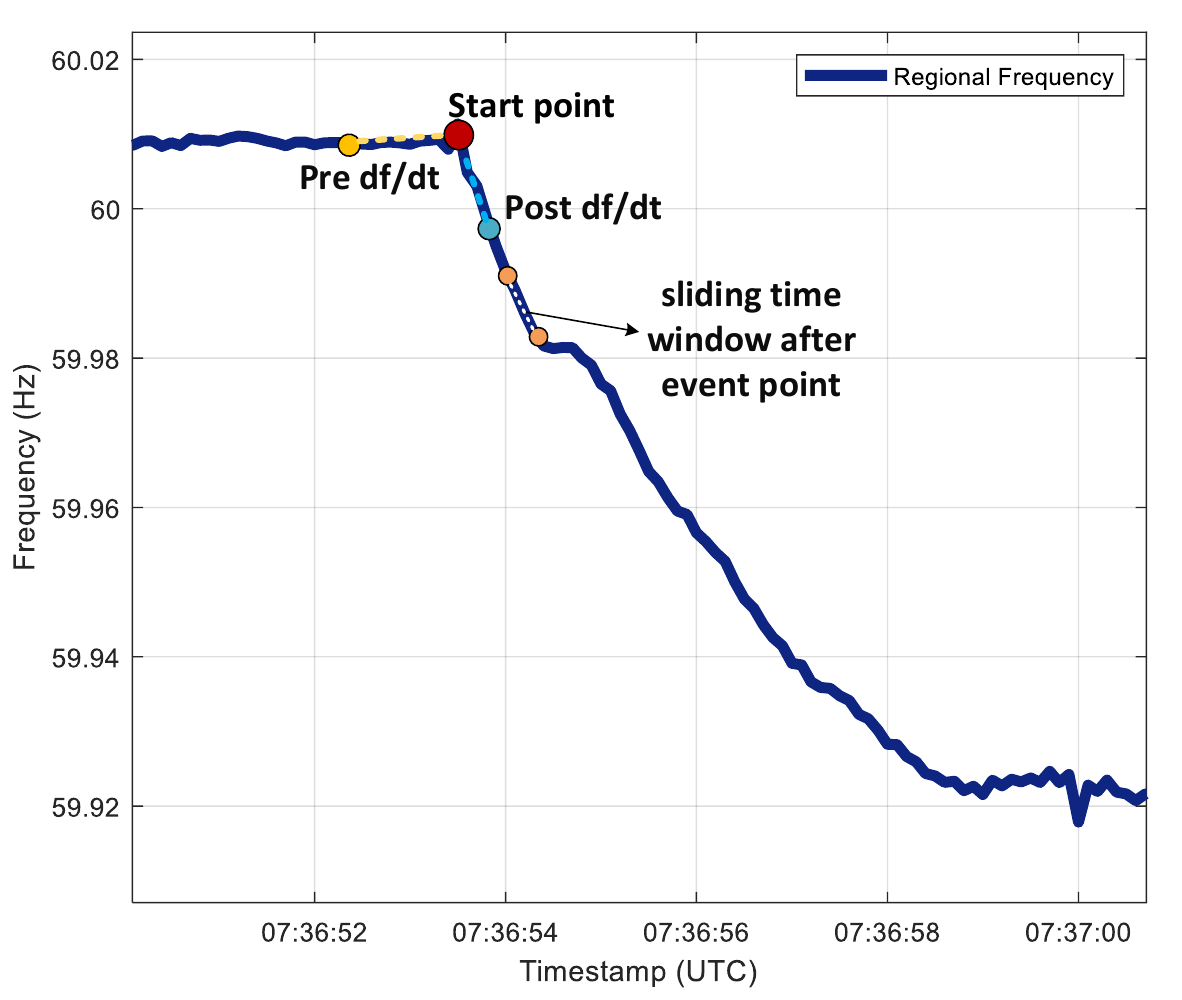}
\caption{Event detection and regional RoCoF estimation for a real event occurred in CAISO.}
\label{fig:rocof}
\end{figure}

Once the regional frequency trace is estimated and the event-point is detected, the RoCoF is computed using a sliding-window method designed for event-driven analysis. Specifically, we apply a 0.1-second moving window across the first 0.5 seconds following the disturbance onset, indicated by the dotted white line between the two yellow points in Fig.~\ref{fig:rocof}. This captures the most critical portion of the system’s inertial response.

The peak RoCoF within this interval is selected as the representative value for inertia estimation. This approach ensures that short-duration transients are accurately captured without being smoothed out by wider averaging.
By operating within the 0.5 sec time window recommended by NERC \cite{NERC}, the method balances noise suppression with temporal precision and reliable inertia assessment at the regional scale.

\subsection{Inertia Computation}

The final step involves calculating inertia using the event-specific $\Delta P$ from NERC-confirmed generation trips. Only events located within the region are selected to ensure regional relevance. Using~\eqref{eqn:swing}, inertia is expressed in MVA·s, representing its energy-equivalent magnitude in the system’s dynamic response.

\section{Advanced Metrics for Regional Dynamic Characterization}

To better understand and quantify regional frequency dynamics following large disturbances, we introduce a set of advanced metrics that highlight both the local and system-wide behavior. Together, they describe (i) how fast frequency initially deviates, (ii) how that deviation varies spatially, (iii) how quickly inertial power arrives from outside CAISO, and (iv) how much inertial energy CAISO itself contributes relative to the entire grid.  

\subsection{Interconnection RoCoF}
The maximum magnitude of the initial RoCoF of the U.S. Western Interconnection (WECC) is selected as the interconnection RoCoF. It is computed by first constructing a system-wide reference frequency using the median of all available FDR measurements, following the method in~\cite{dulal2024inertia}.  
This median-based approach provides a robust estimate of the central frequency trajectory, mitigating the influence of local anomalies or measurement noise.
As a system-level indicator, the interconnection RoCoF reflects the combined inertial response of the whole WECC footprint and serves as a baseline for evaluating regional and local deviations.  

\subsection{Regional RoCoF} 
As detailed in Section III.C, we compute regional RoCoF over the first 0.5 sec following the disturbance. A higher regional RoCoF implies that the region is more susceptible to rapid frequency deviations, often due to high IBR penetration or insufficient local inertia.  
It serves as a key indicator of regional stability and helps differentiate regional response characteristics from the broader interconnection behavior.

\subsection{Local RoCoF}
Local RoCoF focuses on the point-wise or individual FDR-level measurement of RoCoF. It helps to capture spatial variability within a region like CAISO, especially where pockets of high IBR deployment may yield sharper local frequency slopes. Comparing local RoCoFs within the region helps identify weak nodes or sub-regions requiring fast frequency response support.

\subsection{Inertial Support Arrival Time}
This metric quantifies the delay between the onset of frequency decline in the CAISO region and the initiation of frequency decline observed at the broader interconnection level. It is computed by measuring the time difference between the initial frequency drop in CAISO region and the subsequent frequency drop across the entire interconnection. A shorter inertial support arrival time indicates a stronger dynamic coupling between CAISO and the broader Western Interconnection, enabling quicker inertial support during disturbances.

\subsection{Regional-to-System Inertia Ratio ($H_{region}/H_{intercon}$)}
This ratio compares the estimated inertia of the region under study (CAISO) to that of the entire interconnection (WECC). This normalized metric provides insight into the region’s capability to independently manage frequency stability following disturbances. A lower ratio indicates greater dependency on external inertial support, while a higher ratio suggests stronger regional self-sufficiency in mitigating frequency excursions.

\section{Results and Discussion}
With the aforementioned metrics established, we analyze seven confirmed grid disturbance events that occurred in the CAISO region between 2013 and 2024, as verified by NERC. For each event, we estimate RoCoF and inertia at interconnection, regional, and local levels, calculate inertial support arrival times, and evaluate the regional-to-interconnection inertia ratio. Table~\ref{tab:caiso_metrics} summarizes the detailed metrics for each event.
\begin{table*}[h]
\centering
\caption{CAISO Frequency and Inertia Metrics During Major Disturbances}
\resizebox{\textwidth}{!}{
\begin{tabular}{cccccccc}
\toprule
\textbf{Event Date (PST/PDT)} & 7/10/2013 9:49 & 2/2/2014 12:29	& 10/9/2017 12:14 & 12/1/2018 11:06	&  10/15/2021 17:49	& 4/6/2022 15:05 & 8/31/2024 0:36 \\
\midrule
Power mismatch (MW) & 1130 & 1450 & 973 & 1114 & 988 & 794 & 771  \\
Interconnection max RoCoF (mHz/s) & 42 & 58 & 104 & 39 & 34 & 44 & 27  \\
Regional RoCoF (mHz/s) & 207 & 261 & 422 & 227 & 114 & 157 & 65  \\
Local RoCoF (mHz/s) & 307 & 435 & 1139 & 383 & 178 & 335 & 133  \\
$H_{\text{intercon}}$ (MVA$\cdot$s) & 815,000 & 502,000 & 280,000 & 860,000 & 864,000 & 536,000 & 855,083  \\
$H_{\text{region}}$ (MVA$\cdot$s) & 164,000 & 167,000 & 69,200 & 147,000 & 261,000 & 152,000 & 355,850 \\
$H_{\text{local}}$ (MVA$\cdot$s) & 110,000 & 100,000 & 25,600 & 85,000 & 167,000 & 71,100 & 173,910 \\
Inertial support arrival time (s) & 0.2 & 0.2 & 0.2 & 0.2 & 0.0 & 0.1 & 0.2  \\
$H_{\text{region}} / H_{\text{intercon}}$ (\%) & 20.1\% & 33.3\% & 24.7\% & 17.1\% & 30.2\% & 28.3\% & 41.6\%  \\
\bottomrule
\end{tabular}}
\label{tab:caiso_metrics}
\end{table*}

Across all events, the CAISO region experiences high regional RoCoF, with values up to six times greater than the interconnection-wide RoCoF. This highlights the region's vulnerability due to its elevated share of IBRs. However, the average inertial support arrival time from neighboring regions remains consistently near 0.2 seconds, reflecting strong electrical coupling between CAISO and the wider Western Interconnection. This coupling underscores the critical role of CAISO’s robust transmission network and favorable grid topology in mitigating regional frequency excursions.

\begin{figure}[ht]
\centering
\includegraphics[width=\linewidth]{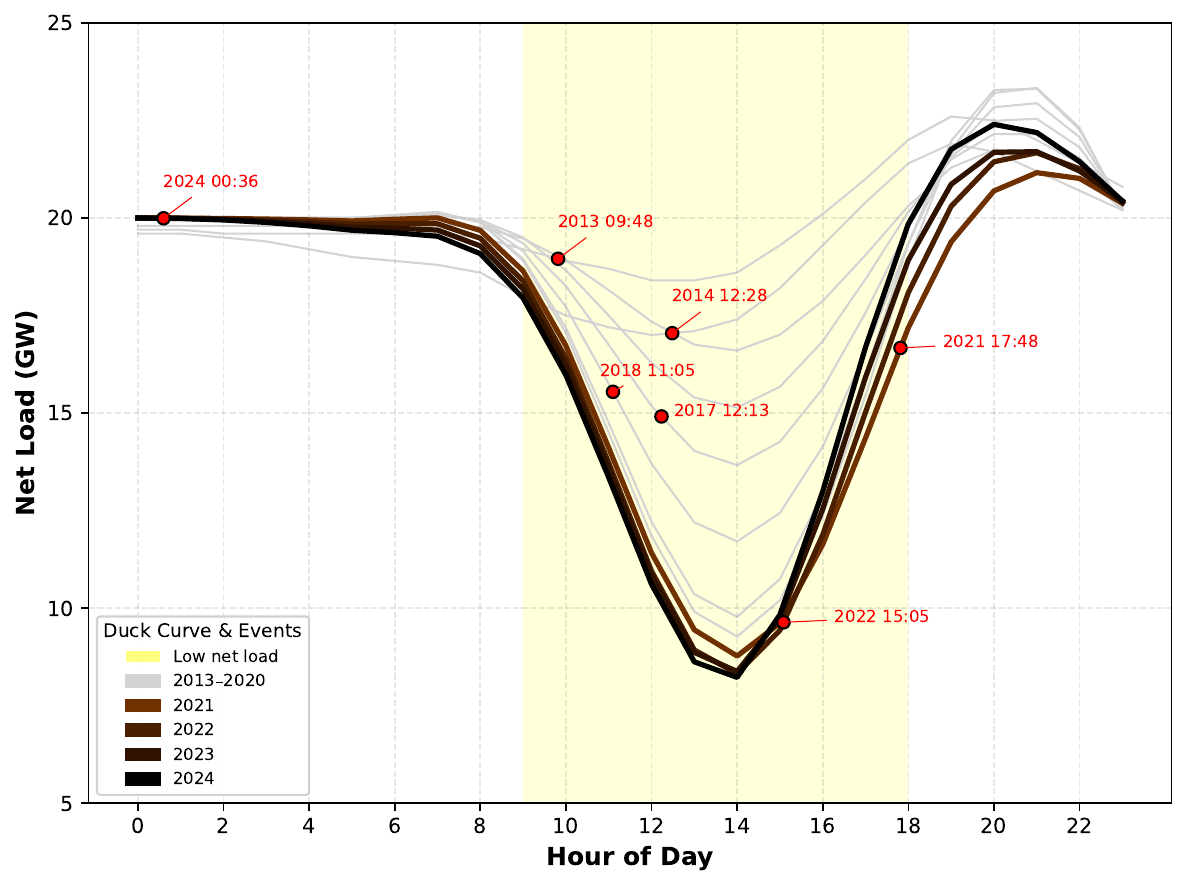}
\caption{CAISO duck curve with confirmed events marked with red dots. Curves are conceptual and reflect typical net-load patterns from 2013 to 2024 based on public CAISO and EIA data \cite{EIAduckcurve}.}
\label{fig:duckcurve}
\end{figure}

An analysis of the temporal distribution of events, as illustrated in Fig.~\ref{fig:duckcurve}, reveals that most disturbances occurred during the daytime window. The clustering of events typically occurs in the ``belly" of the duck curve, where net load is at its lowest, as highlighted in yellow in the figure. These periods coincide with significantly reduced net load—defined as system demand minus renewable generation—and are characterized by diminished synchronous inertia. This reduction in inertia is primarily due to high solar generation and curtailed conventional generation. In contrast, the event recorded on August 31, 2024, occurred outside this low net load window and exhibited a higher regional-to-interconnection inertia ratio. This observation illustrates the influence of net load dynamics on grid stability and available inertia. 

The duck curve began to deepen in 2014, driven by the rapid growth of solar generation. This trend coincided with the lowest levels of regional inertia observed between 2014 and 2018, particularly during midday events when solar output was at its peak. During this period, both absolute and relative inertia values declined (see Table~\ref{tab:caiso_metrics}), contributing to elevated RoCoF levels and reduced frequency resilience.

Following years of declining inertia, CAISO began to recover in 2019, driven by advancements in battery energy storage systems (BESS) and grid-forming technologies. The deployment of front-of-the-meter BESS capacity increased significantly, from 0.2 GW in 2018 to 13 GW by 2024~\cite{CAISOBESS}. These batteries play a dual role in grid stability by absorbing surplus solar generation during midday and discharging energy to meet the steep ramp in demand during evening hours. In 2024, they supplied an average of 8.6\% of balancing-area energy between 17:00 and 21:00, with momentary injections up to 6 GW.

\sloppy
Concurrently, CAISO adopted grid-forming inverter technologies and improved ride-through requirements, aligning with NERC’s bulk power system (BPS)-connected BESS guidelines and Federal Energy Regulatory Commission (FERC) Orders 841/842 on storage integration and frequency response \cite{NERCBESS}. Events in 2021 and 2022, which occurred during afternoon and evening hours, respectively, exhibited higher regional inertia to interconnection ratios, suggesting a growing contribution from synthetic and fast-frequency-responsive resources in CAISO. These developments underscore the importance of strategic resource planning and operational practices to improve resilience during periods of high renewable generation.

s

\section{Conclusion}
This paper presents a measurement-based framework for estimating regional inertia using real disturbance data from FNET/GridEye. It addresses the need to assess inertia beyond the interconnection level. By analyzing seven major events in CAISO, we identify distinct regional inertia characteristics, including elevated RoCoF values, earlier inertial response arrival times, and lower regional inertia during midday disturbances. These variations often correspond to reduced net load resulting from high solar output and evolving system dynamics. Our results show a declining inertia trend through 2018, followed by a recovery phase likely supported by growing deployment of front-of-the-meter BESS and grid-forming controls. Despite these improvements, spatial imbalances persist. These findings reinforce the importance of region-specific inertia monitoring, dynamic situational awareness, and adaptive control strategies to maintain frequency stability during the grid transition. Future research will focus on applying this framework to other regions with comparable challenges and enhancing the metrics to reflect the changing dynamics of low-inertia systems.

\section*{Acknowledgment}
This material is based upon work supported by the US Department of Energy, Office of Electricity (OE) under contract DE-AC05-00OR22725.

\bibliographystyle{IEEEtran}
\bibliography{references}

\end{document}